\documentclass[epj]{svjour}
\usepackage{longtable}
\usepackage{epsfig,graphicx}
\usepackage{lscape}

\begin{document}

 \title{Rejection of randomly coinciding events in ZnMoO$_4$ scintillating bolometers}

 \author{D.M.~Chernyak\inst{1,}\inst{2}  \and F.A.~Danevich\inst{1} \and A.~Giuliani\inst{2,}\inst{3}\thanks{\emph{Corresponding author;} e-mail: Andrea.Giuliani@csnsm.in2p3.fr (A.~Giuliani).} \and
M.~Mancuso\inst{2,}\inst{3} \and C.~Nones\inst{4} \and
E.~Olivieri\inst{2} \and M.~Tenconi\inst{2}\and
V.I.~Tretyak\inst{1}}

 \institute{Institute for Nuclear Research, MSP 03680 Kyiv, Ukraine
 \and Centre de Sciences Nucl\'{e}aires et de Sciences de la Mati\`{e}re, 91405 Orsay, France
 \and Dipartimento di Scienza e Alta Tecnologia dell'Universit\`{a} dell'Insubria, I-22100 Como, Italy
 \and Service de Physique des Particules, CEA-Saclay, F-91191 Gif sur Yvette, France}

\date{Received: date / Revised version: date}

\abstract{Random coincidence of events (particularly from
two neutrino double beta decay) could be one of the main sources of background
in the search for neutrinoless double beta decay with cryogenic bolometers due to their poor
time resolution. Pulse-shape discrimination by using front edge
analysis, mean-time and $\chi^2$ methods was applied to
discriminate randomly coinciding events in ZnMoO$_4$ cryogenic
scintillating bolometers. These events can be effectively
rejected at the level of 99\% by the analysis of the heat signals with rise-time of about 14 ms and signal-to-noise ratio of 900, and at the level of 92\% by the analysis of the light signals with rise-time of about 3 ms and signal-to-noise ratio of 30, under the requirement to detect 95\% of single events. These rejection efficiencies are compatible with extremely low background levels in the region of interest of neutrinoless double beta decay of $^{100}$Mo for enriched ZnMoO$_4$ detectors, of the order of $10^{-4}$~counts/(y keV kg). Pulse-shape parameters have been chosen on the basis of the performance of a real massive ZnMoO$_4$ scintillating bolometer. Importance of the signal-to-noise ratio, correct finding of the signal start and
choice of an appropriate sampling frequency are discussed.}

\titlerunning{Rejection of randomly coinciding events in scintillating bolometers}
\authorrunning{D.M.~Chernyak {\it et al}.,}
\maketitle

\section{Introduction}

Observation of neutrinoless double beta ($0\nu2\beta$) decay would
imply the violation of lepton number conservation and definitely
new physics beyond the Standard Model, establishing the Majorana
nature of neutrino
\cite{Avignone:2008,Rodejohann:2011,Elliott:2012,Vergados:2012,Giuliani:2012a}.
Cryogenic scintillating bolometers look the most promising
detectors to search for this extremely rare process in a few
theoretically favourable nuclei
\cite{Pirro:2006,Arnaboldi:2010,Lee:2011,Giuliani:2012b,Beeman:2012a,Beeman:2012b,Beeman:2013}.
Zinc molybdate (ZnMoO$_4$) is one of the most promising materials
to search for $0\nu2\beta$ decay thanks to the absence of
long-lived radioactive isotopes of constituting elements,
the comparatively high percentage of molybdenum and the recently developed
technique of growing large high quality radiopure ZnMoO$_4$ crystal
scintillators
\cite{Beeman:2012a,Beeman:2012b,Gironi:2010,Beeman:2012c,Beeman:2012d,Chernyak:2013}.

However, a disadvantage of the low temperature bolometers is their
poor time resolution, which can lead to a significant background
at the energy $Q_{2\beta}$ due to random coincidences of signals,
especially of the unavoidable two-neutrino $2\beta$ decay events
\cite{Chernyak:2012}. This issue is particularly relevant for the
experiments aiming at searching for $0\nu2\beta$ decay of $^{100}$Mo,
because of the short half-life of $^{100}$Mo in comparison
to the two neutrino double beta ($2\nu2\beta$) decay
$T_{1/2}=7.1\times10^{18}$ yr \cite{Arnold:2005}. Counting rate of
two randomly coincident $2\nu2\beta$ events in cryogenic
Zn$^{100}$MoO$_4$ detectors is expected to be on the level of
$2.9\times10^{-4}$ counts / (keV$\times$kg$\times$yr) at the
$Q_{2\beta}$ energy (for 100 cm$^3$ crystals, under a condition
that two events shifted in time in 1 ms can be resolved), meaning
that randomly coincident $2\nu2\beta$ decays can be even a main
source of background in a future large scale high radiopurity
experiment \cite{Chernyak:2012}.

This work describes the development of pulse shape discrimination
techniques to reject randomly coinciding events in ZnMoO$_4$
cryogenic scintillating bolometers.

\section{Randomly coinciding events in cryogenic bolometers}
\label{sec:gen}

\begin{figure}[htbp]
\begin{center}
\mbox{\epsfig{figure=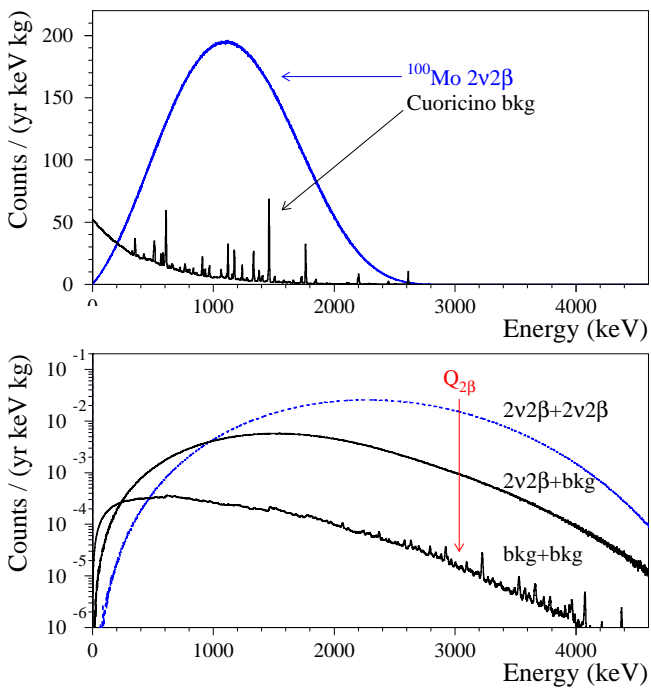,height=9.0cm}}
\caption{Distribution for the sum of energies of two electrons
emitted in $2\nu2\beta$ decay of $^{100}$Mo and a model of the background energy spectrum from external gamma quanta~\cite{Andreotti:2011} (upper panel), and Monte Carlo simulated energy spectra of two randomly coincident $2\nu2\beta$ events, coincident $2\nu2\beta$ with external gamma events, and randomly coincident external gamma events (lower panel).}
 \label{fig:RC-2n2b-ext}
\end{center}
\end{figure}

The energy distribution of the randomly coinciding (rc)
$2\nu2\beta$ events was built using the approach described in
\cite{Chernyak:2012}, under the assumption that two events are not
resolved in the time interval 45 ms.\footnote{This time interval is related to typical rise-times observed in large mass bolometers, like those operated in the Cuoricino experiment~\cite{Andreotti:2011}.} The energy spectra of
$10^9$ Monte Carlo generated events of $^{100}$Mo $2\nu2\beta$
decay and of two randomly coinciding $2\nu2\beta$ events are
presented in Fig.~\ref{fig:RC-2n2b-ext}. The rate for the rc
events is calculated as $I_{rc} = \tau \cdot I_0^2$, where $I_0 =
\ln 2~N/T_{1/2}$, $N$ is the number of $2\beta$ decaying nuclei in
Zn$^{100}$MoO$_4$ crystal (with 100\% enrichment by $^{100}$Mo) of
typical size of $\oslash6\times4$ cm ($N = 1.28\times10^{24}$),
$\tau = 45$ ms, $T_{1/2}=7.1\times10^{18}$ yr.

\nopagebreak

In addition to the $2\nu2\beta$ decay, there could be other
background sources contributing to the background in the region of
interest due to random coincidence. We have estimated a possible 
contribution of the external gamma background using the level of
background already achieved in the Cuoricino detector
\cite{Andreotti:2011}. A simplified model of the Cuoricino
background (taken from \cite{Andreotti:2011} in the energy
interval 300 -- 2620 keV, while exponentially extrapolated below
300 keV and equated to 0 above 2620 keV) and a Monte Carlo
simulated energy spectrum of two randomly coincident background
events are presented in Fig.~\ref{fig:RC-2n2b-ext}.

We have also simulated coincidences of the Cuoricino background
with the $2\nu2\beta$ decay of $^{100}$Mo. One can see that the main
contribution to background (assuming radiopure Zn$^{100}$MoO$_4$
crystal scintillators and a cryostat with a level of radioactive
contamination similar to that of the Cuoricino set-up) is expected from the
$2\nu2\beta$ decay of $^{100}$Mo. The total counting rate due to
the random coincidences of $2\nu2\beta$ decay events and external
gamma events in the region of interest is estimated as
$\approx0.016$ counts/(year keV
 kg) for a detector time resolution of 45 ms.

\section{Generation of randomly coinciding signals}

Sets of single and randomly coincident signals were generated by
using pulse profiles and noise baselines accumulated with a real 0.3 kg
ZnMoO$_4$ crystal scintillator operated as a cryogenic
scintillating bolometer with a Ge light detector~\cite{LD-tenconi} in Centre de Sciences Nucl\'{e}aires et de
Sciences de la Mati\`{e}re (Orsay, France). Two measurements have been taken into account, the first one with a sampling rate of
5 kSPS (kilosamples per second) both for the light and heat channels, and the second one with the sampling rate 1.9841 kSPS for the both channels. Ten thousand of base-line samples were selected in all the cases.

The pulse profiles of heat and light signals of the detectors (sum
of a few hundred pulse samples produced mainly by cosmic rays with
energy of a few MeV) were obtained by fit with the following
phenomenological function:
\begin{equation}
 f_S(t)=A\cdot(e^{-t/\tau_1}+e^{-t/\tau_2}-e^{-t/\tau_3}-e^{-t/\tau_4}),
\label{eq:shape}
\end{equation}
where $A$ is the amplitude,  $\tau_1$, $\tau_2$, $\tau_3$ and
$\tau_4$ are the time constants.

To generate randomly coinciding signals in the region of
the $Q_{2\beta}$ value of $^{100}$Mo, the amplitude of the first pulse
$A_1$ was obtained by sampling the $2\nu2\beta$ distribution for
$^{100}$Mo, while the amplitude of the second pulse was chosen as
$A_2=Q_{2\beta} - A_1 + \Delta E$, where $\Delta E$ is a random
component in the energy interval $[-5,+5]$ keV (which is a typical
energy resolution of a bolometer).

Ten thousand coinciding signals were randomly generated in the
time interval from $0$ to $3.3\cdot\tau_R$ $(\Delta
t=[0,3.3\cdot\tau_R]$, see Section~\ref{sec:mtm}), where $\tau_R$ is the rise-time of the signals
(defined here as the time to increase the pulse amplitude from 10\%
to 90\% of its maximum). As it will be demonstrated in the section
4.1, the rejection efficiency of randomly coinciding signals (RE,
defined as the part of the pile-up events rejected by pulse-shape
discrimination) reaches almost its maximal value when the time
interval of consideration exceeds $(3-4)\tau_R$. Ten thousand of
single signals were also generated.

\begin{figure}[htbp]
\begin{center}
\mbox{\epsfig{figure=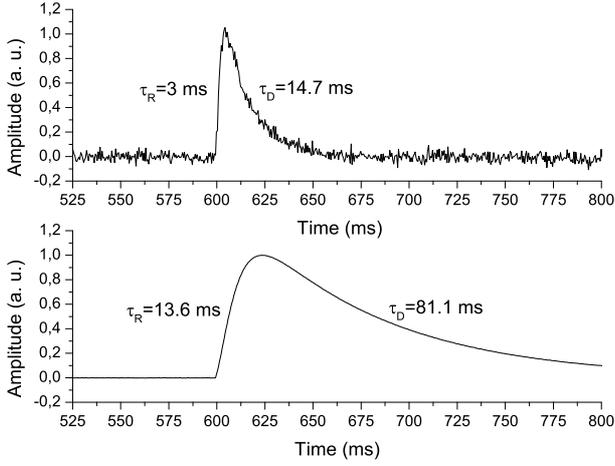,height=6.5cm}} \caption{Examples of
generated light (upper panel) and heat (lower panel) pulses.
$\tau_R$ and $\tau_D$ denote rise- and decay-times, respectively.}
\label{signals}
\end{center}
\end{figure}

A signal-to-noise ratio (defined as the ratio of the maximum signal
amplitude to the standard deviation of the noise baseline) was
taken 30 for light signals and 900 for heat signals. These values
are typical for ZnMoO$_4$ scintillating bolometers. Examples of
the generated heat and light single pulses are presented in Fig.
\ref{signals}.

\nopagebreak

\section{Methods of pulse-shape discrimination}

We have applied three techniques to discriminate randomly coincident events: mean-time method, $\chi^2$ approach, and front edge analysis. We demanded a 95\% efficiency in accepting single signals. For an efficient discrimination, it is very important to develop a method for a good determination of the pulse start position.

\subsection{Reconstruction of the time origin of the events}
\label{sec:origin}

The following procedure was used to reconstruct the time origin of each signal:
\begin{enumerate}
\item Preliminary search for the presence of a signal by a very simple algorithm, which searches for a channel where the signal amplitude exceeds a certain level (typically about one third of the signal maximum value);
\item Summation of the data over a certain number of channels (typically over 2-6 channels for the light signals, depending on the time structure of signal and noise data; this procedure was not used for the heat signals);
\item Calculation of the standard deviation of the baseline fluctuations;
\item Search for the pulse start under the request that the signal exceeds a certain number of standard deviations of the baseline and -- in a case of heat signals -- the amplitude in the next several channels increases channel by channel.
\end{enumerate}

The algorithm was optimized for each data set, taking into account the sampling rate, the time properties of the signals and noise, and the signal-to-noise ratio.

\subsection{Mean-time method}
\label{sec:mtm}

The following formula was applied to calculate the parameter
$\langle t \rangle$ (mean-time) for each pulse $f(t_k)$:
\begin{equation}
 \langle t \rangle =\sum f(t_k)t_k/\sum f(t_k),
\end{equation}
where the sum is over time channels $k$, starting from the origin
of a pulse and up to a certain time.

\begin{figure}[htbp]
\begin{center}
\mbox{\epsfig{figure=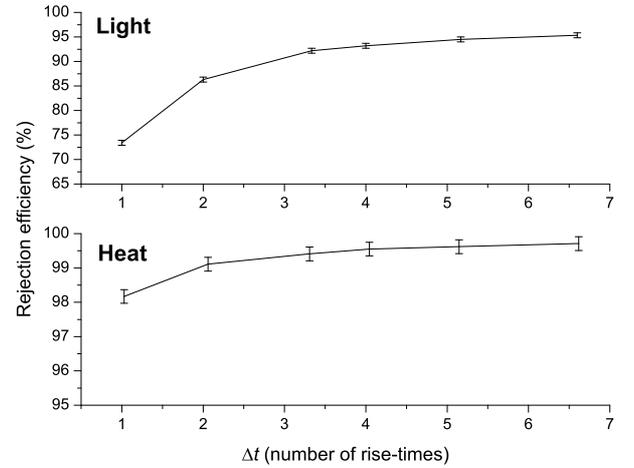,height=6.5cm}} \caption{Dependence
of the rejection efficiency (by using the mean-time method) for
heat and light channels on the time interval $\Delta t$ where the
randomly coinciding signals were generated.}
\label{t-int}
\end{center}
\end{figure}

As a first step we have chosen the time interval $\Delta t$ to
analyze efficiency of the pulse-shape discrimination. Six sets of
single and randomly coinciding light (with $\tau_R=3$ ms) and heat
signals (with $\tau_R=13.6$ ms) were generated in the time
intervals ranging from 0 to a maximum value, varying from 1 to about 6 pulse rise-times. The results of this analysis are presented in Fig. \ref{t-int}. The uncertainties of
the rejection efficiency were estimated by analysis of three sets
of data generated using three sets of different noise baseline
profiles (about 3300 profiles in the each set). One can see that
the rejection efficiency of randomly coinciding signals reaches
its maximal value when the time interval $\Delta t$ is larger than
$(3-4)\tau_R$. All the further analysis was done by using data
generated in the time interval $\Delta t=[0,3.3\tau_R]$. 

\nopagebreak

\begin{figure}[htbp]
\begin{center}
\mbox{\epsfig{figure=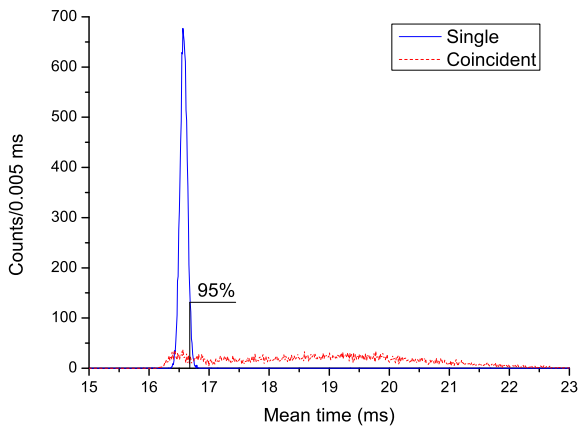,height=6.5cm}}
\caption{Distribution of the mean-time parameter for single and
coincident light pulses with a rise-time 3 ms. Rejection
efficiency of coincident pulses is 92.2\%. The events left from
line are accepted as single events (95\% of single events). 7.8\%
of pile-up events moves to the ``single'' event region due to
incorrect start finding and / or too small time difference between
coinciding signals.}
 \label{mean-time}
\end{center}
\end{figure}

A typical distributions of the mean time parameters for single and
pile-up events are presented in Fig. \ref{mean-time}. The
rejection efficiency of randomly coinciding pulses, under the
requirement to detect 95\% of single events, is 92.2\%.

\nopagebreak

\begin{figure}[htbp]
\begin{center}
\mbox{\epsfig{figure=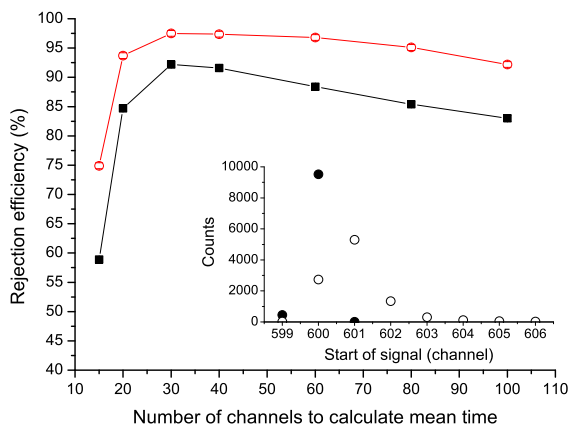,height=6.5cm}} \caption{Dependence
of the rejection efficiency of the mean-time method on the number
of channels used to calculate the parameter $\langle t \rangle$. The
analysis was performed for the light signals with 3 ms rise-time.
The rejection efficiency of randomly coinciding pulses is 92.2\% for
the cases when the start of the signals was found by our algorithm
(squares), and 97.5\% using the known start position (circles). One
channel is 0.504 ms. (Inset) Distribution of start positions for
the sets of single (filled circles) and randomly coinciding (open
circles) events.} \label{RE-chan-N}
\end{center}
\end{figure}

One could expect that the rejection efficiency of pulse-shape discrimination depends on the choice of the time interval used to
calculate a discrimination parameter. For instance, in Fig.
\ref{RE-chan-N} the results of the mean-time method optimization
are presented. The rejection efficiency has a maximum when the mean-time
parameter is calculated from the signal origin to the 30th channel
which approximately corresponds to $\sim \tau_D$. All the discrimination
methods were optimized in a similar way.

\nopagebreak

The dependences presented in Fig. \ref{RE-chan-N} demonstrate also
the importance to optimize an algorithm to find the start of the signals (see Section~\ref{sec:origin}),
particularly for the light pulses with comparatively low
signal-to-noise ratio. Rejection efficiency is substantially
higher when the start position of each pulse is known from
the generation algorithm. The distributions of start positions for
the sets of single and randomly coinciding events shown in Inset
of Fig. \ref{RE-chan-N} demonstrate that it is more
problematic to correctly find the start position of randomly coinciding signals than that of single events.

\subsection{$\chi^2$ method}

The approach is based on the calculation of the $\chi^2$ parameter
defined as
\begin{equation}
 \chi^2 =\sum (f(t_k)-f_S(t_k))^2,
\end{equation}
where the sum is over time channels $k$, starting from the origin
of pulse and up to a certain time, and $f_S(t)$ is defined by Eq.~(\ref{eq:shape}). The number of channels to
calculate the $\chi^2$ has been optimized to reach a maximal
rejection efficiency.

\subsection{Front edge analysis}

The front edge parameter can be defined as the time between two
points on the pulse front edge with amplitudes $Y_1$\% and $Y_2$\%
of the pulse amplitude. The parameters $Y_1$ and $Y_2$ should be
optimized to provide maximal rejection efficiency. For instance,
the highest rejection efficiency for heat pulses with
$\tau_R=13.6$ ms was reached with the front edge parameter
determined as time between the signal origin and the time where
the signal amplitude is $Y_2=90\%$ of its maximum (RE = 98.4\%).

However, the rejection efficiency of the front edge method is
limited due to the fraction of randomly coinciding events
with a small first (with the amplitude $A_1$ below $Y_1$) or
second pulse (with a low amplitude, and appearing well after the
first signal maximum).

\section{Results and discussion}

\begin{figure}[htbp]
\begin{center}
\mbox{\epsfig{figure=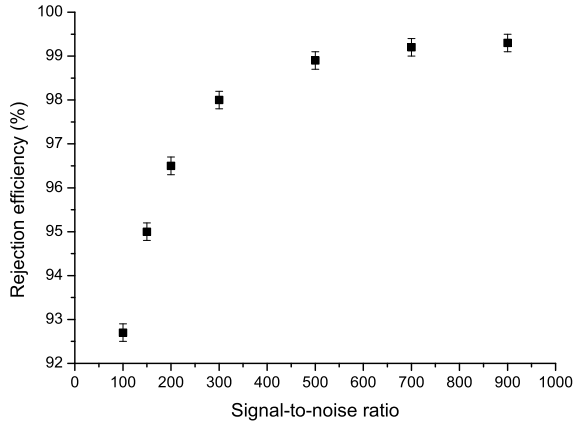,height=6.5cm}} \caption{Dependence
of the rejection efficiency (using the mean-time method) on the
signal-to-noise ratio for the heat channel. }
 \label{signal-noise}
\end{center}
\end{figure}

The methods of pulse-shape discrimination are compared in Table
\ref{tb:summary}. The data were obtained with start positions of
the signals found by our algorithms, and using {\it a priori} known
signal start positions from the generation procedure (to estimate
a maximum achievable efficiency). All the methods give a $86\%-92\%$
rejection efficiency by using the light signals with a rise-time of 3 ms
and $98\%-99\%$ for the much slower heat signals with a rise-time
of 13.6 ms. One can conclude that the signal-to-noise ratio (set to 30 for
the light and to 900 for the heat signals at the energy $Q_{2\beta}$, as observed in real bolometers) plays a crucial role in the pulse-shape discrimination of randomly coinciding events in
cryogenic bolometers. Analysis of signals with lower level of
noise allows to reach much higher rejection efficiency even with
slower heat signals. Dependence of the rejection efficiency (by
using the mean-time method) on the signal-to-noise ratio for heat
signals confirms the assumption (see Fig. \ref{signal-noise}).

\begin{table}[htbp]
\caption{Rejection efficiency of randomly coinciding $2\nu2\beta$
events by pulse-shape discrimination of light and heat signals for
the two conditions of the signal start determination, i.e. (i) start of the signals known from the generation procedure, (ii) and start position found by the pulse profile analysis.}
\begin{center}
\begin{tabular}{|lllll|}
\hline
 Channel,       & Start     & Mean-time     & Front edge    & $\chi^2$ method, \\
 rise-time      & position  & method, \%    & analysis, \%  & \%  \\
  ~             &  ~        & ~             & ~             & ~ \\
 \hline
 Light,         & Known     & $97.5\pm0.5$  & $96.4\pm0.5$  & $97.4\pm0.5$ \\
 3 ms           & Found     & $92.2\pm0.5$  & $88.1\pm0.5$  & $92.3\pm0.5$ \\
 Heat,          & Known     & $99.4\pm0.2$  & $99.4\pm0.2$  & $99.4\pm0.2$ \\
 13.6 ms        & Found     & $99.3\pm0.2$  & $99.3\pm0.2$  & $99.3\pm0.2$ \\
\hline
\end{tabular}
\end{center}
\label{tb:summary}
\end{table}

\nopagebreak

We tried also to analyse a dependence of the rejection efficiency
on the rise-time of light pulses (see Fig.~7). We have assumed that the pulse amplitude does not change by shortening the rise-time. One could expect
that for faster signal any methods should give a higher efficiency
of pulse-shape discrimination. However, the trend of the rejection
efficiency improvement for faster signals is rather weak.
Furthermore, the rejection efficiency for pulses with a rise-time of 2 ms is even worse in comparison to slower signals with
rise-times of 3 ms and 4.5 ms. This feature can be explained by a
rather low sampling rate (1.9841 kSPS) used for the data
acquisition. Indeed, the pulse profiles acquired with this
sampling rate are too discrete: for instance, the front edge of
the signals with the rise-time 2 ms is represented by only 4
points. Such a low discretization even provides difficulties to
set the acceptance factor of single events at a certain level
(95\% in our case), particularly in the front edge analysis.

\begin{figure}[htbp]
\begin{center}
\mbox{\epsfig{figure=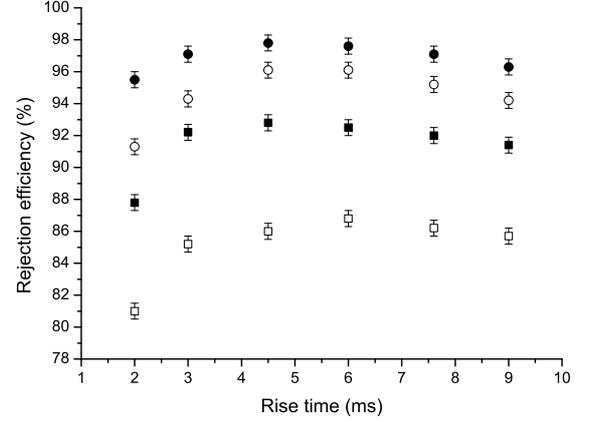,height=6.5cm}} \caption{Dependence
of the rejection efficiency for the light pulses (by using the
mean-time method) on the rise-time, signal-to-noise ratio and data
acquisition sampling rate. The filled squares (circles) represent
the data with a signal-to-noise ratio of 30 (100) acquired with
a sampling rate of 1.9841 kSPS, while the open markers show results
for the same signals acquired with a sampling rate of 0.9921 kSPS.}
\label{RE-light}
\end{center}
\end{figure}

The importance of the data acquisition sampling rate for the pulse-shape analysis was proved by taking the noise baselines data with 2 times lower sampling rate (we have transformed the data by simply averaging two adjacent channels to one). The rejection efficiency decreases in this case, as one can see in Fig. \ref{RE-light}.

\nopagebreak

Following the previous discussion, we will calculate now how our pile-up rejection procedure would improve the background figure estimated at the end of Section~\ref{sec:gen}, and corresponding to $\approx 0.016$ in an enriched Zn$^{100}$MoO$_4$ detector based on a crystal size of $\oslash6\times4$ cm. We remind that this value was obtained assuming that pulses separated by an interval longer than 45 ms would be far enough to be analysed independently and on the contrary all the concomitant pulses within this time interval would give rise to a single amplitude equal to the sum of the individual ones.  We are authorised to abate the resulting background level by 99.3\% in the heat channel and 92.3\% in the light channel, as these are our best pile-up rejection efficiencies with unknown pulse start position (see Table 1).\footnote {We safely assume that the rejection efficiency in the 45 ms time interval is the same as in the shorter $3.3 \cdot \tau_R$ time interval since the rejection efficiency starts to saturate, as it can be appreciated in Fig. 3.} The final estimated background level after pile-up rejection is $\approx 1.1 \times 10^{-4}$ counts/(year keV kg) using pulse-shape discrimination in the heat channel. A higher value is obtained using the light channel because of the worse rejection efficiency.  An improvement of the speed and of the
signal-to-noise ratio in cryogenic bolometers -- both in the heat and light channel -- is an important experimental goal to further enhance the rejection of RC-generated background.

\section{Conclusions}

Random coincidence of events (especially but not only from $2\nu2\beta$ decay)
could be one of the main sources of background in cryogenic
bolometers to search for $0\nu2\beta$ decay because of their poor
time resolution, particularly for $^{100}$Mo due to the short half-life in comparison to that of the two-neutrino mode. However, this
background can be effectively suppressed with the help of
pulse-shape discrimination.

The randomly coinciding $2\nu2\beta$ decay signals were
discriminated with an efficiency at the level of 99\% by applying
the mean-time approach to the heat signals from ZnMoO$_4$
cryogenic bolometer with a rise-time of about 14 ms and a
signal-to-noise ratio of 900, and at the level of 97\% for the light
signals with 3 ms rise-time and signal-to-noise ratio of 30 (however,
the last estimation was obtained for the signals with {\it a
priory} known start position). $\chi^2$ approach provides
comparable rejection efficiencies, while the results of the front
edge analysis are slightly worse due to problems in
discriminating pile-up events when one of the randomly coinciding signals is
too small to be detected by this method.

The signal-to-noise ratio looks the most important feature to
reject randomly coinciding events, particularly in ZnMoO$_4$ due
to the comparatively low light yield, which leads to a rather low
signal-to-noise ratio in the light channel.

Development of algorithms to find the origin of a signal with as high
as possible accuracy is requested to improve the rejection capability
of any pulse-shape discrimination technique. The sampling rate of the
data acquisition should be high enough to provide effective
pulse-shape discrimination of randomly coinciding events. Finally,
any pulse-shape discrimination methods should be optimized taking
into account certain detector performance to reduce the background
effectively.

The analysis performed in this work proves further that the
counting rate due to the random coincidence of events can be
reduced to a level $\approx10^{-4}$ counts/(year keV
 kg), which makes ZnMoO$_4$ cryogenic scintillating bolometers very promising to
search for neutrinoless double beta decay at a level of
sensitivity high enough to probe the inverted hierarchy region of the neutrino mass pattern~\cite{Beeman:2012a}.

\section{Acknowledgements}

This work was supported in part by the project ``Cryogenic detector
to search for neutrinoless double beta decay of molybdenum'' in the
framework of the Programme ``Dnipro'' based on the Ukraine-France
Agreement on Cultural, Scientific and Technological Cooperation.
The study of ZnMoO$_4$ scintillating bolometers is part of the
program of ISOTTA, a project receiving funds from the ASPERA 2nd
Common Call dedicated to R\&D activities.

\end{document}